\documentclass[12pt,onecolumn]{IEEEtran}
\usepackage{amsmath}
\usepackage{amssymb}
\usepackage{mathrsfs}
\usepackage{cite}
\usepackage{epsfig}
\usepackage{epsf}
\usepackage{theorem}
\usepackage{graphics}
\newenvironment{Proof}{\textit{Proof:}}{$\blacksquare$}

\newtheorem{thm}{Theorem}

\newtheorem{lemma}{Lemma}

\newtheorem{cor}{Corollary}

\begin{document}
\pagenumbering{arabic} \setcounter{page}{1}

\title{On The Limitations of The Naive Lattice Decoding}
\author{
Mahmoud Taherzadeh and Amir K. Khandani\\
\small Coding \& Signal Transmission
Laboratory(www.cst.uwaterloo.ca)\\ \small Dept. of Elec. and Comp.
Eng., University of Waterloo, Waterloo, ON, Canada, N2L 3G1 \\\small
e-mail: \{taherzad, khandani\}@cst.uwaterloo.ca, Tel:
519-8848552, Fax: 519-8884338} \maketitle

\begin{abstract}
In this paper, the inherent drawbacks of the naive lattice decoding for MIMO fading systems is investigated. We show that using
the naive lattice decoding for MIMO systems has considerable
deficiencies in terms of the rate-diversity trade-off. Unlike the case of maximum-likelihood decoding, in this case, even the perfect lattice space-time codes which have the non-vanishing determinant property can not achieve the optimal rate-diversity trade-off. Indeed, we show that in the case of naive lattice decoding, when we fix the underlying lattice, all the codes based on full-rate lattices have the same rate-diversity trade-off as V-BLAST. Also, we drive a lower bound on the symbol error probability of the naive lattice
decoding for the fixed-rate MIMO systems (with equal numbers of
receive and transmit antennas). This bound shows that
asymptotically, the naive lattice decoding has an unbounded loss in
terms of the required SNR, compared to the maximum likelihood decoding\footnote{Financial support provided by Nortel and the corresponding matching funds by the Natural Sciences
and Engineering Research Council of Canada (NSERC), and Ontario Centres of Excellence (OCE) are
gratefully acknowledged.}.

\end{abstract}

\section{Introduction}

In recent years, there has been extensive research on designing practical encoding/decoding schemes to approach theoretical limits of MIMO fading systems. The optimal rate-diversity trade-off \cite{ZT2003} is considered as an important theoretical benchmark for practical systems. For the encoding part, recently, several lattice codes are introduced which have the non-vanishing determinant property and achieve the optimal trade-off, conditioned on using the exact maximum-likelihood decoding \cite{KumarRDT1} \cite{BelfioreRDT} \cite{KumarRDT2}. The lattice structure of these codes facilitates the encoding. For the decoding part, various lattice decoders, including the sphere decoder and the lattice-reduction-aided decoder are presented in the literature 
\cite{damen2000} \cite{fischer2003}. To achieve the
exact maximum likelihood performance, we need to find the closest
point of the lattice inside the constellation region, which can be
much more complex than finding the closest point in an infinite
lattice. To avoid this complexity, one can perform the traditional
lattice decoding (for the infinite lattice) and then, discard the
out-of-region points. This approach is called Naive Lattice
Decoding (NLD).

In \cite{taherzad-IT2}, the authors have shown that
this sub-optimum decoding (and even its lattice-reduction-aided approximation) still achieve the maximum receive diversity in
the fixed-rate MIMO systems. Achieving the optimal receive diversity by a low decoding complexity makes lattice-reduction-aided decoding (using the LLL reduction) an attractive choice for different applications. Nonetheless, this work shows that concerning rate-diversity trade-off, the optimality can not be achieved by the naive-lattice decoding or its approximations.

In \cite{DamenElgamalCaire}, using a probabilistic method, a lower bound on the best achievable trade-off, using the naive lattice decoding, is presented. 
In this paper, we present an upper bound on the performance of the naive lattice decoding for codes based on full-rate lattices. We show that
NLD can not achieve the optimum rate-diversity trade-off. Also, for the special case of equal number of transmit and receive antennas, we show that even the best full-rate lattice codes (including perfect space-time codes such as the Golden code \cite{BelfioreRDT}) can not perform better than the simple V-BLAST (if we use the naive lattice decoding at the receiver). It should be noted that in this paper, we have assumed that the underlying lattice is fixed for different rates and SNR values (e.g. lattice codes introduced in \cite{KumarRDT1} \cite{BelfioreRDT} \cite{KumarRDT2}). If we relax this restriction, there can exist a family of lattice codes (based on different lattice structures for different rates and SNR values) which achieves the optimum tradeoff using the naive lattice decoding \cite{ElGamal2007}.

In section IV, we complement the result of \cite{taherzad-IT2} by showing that for the special case of equal number of transmit and receive antennas, although the naive lattice decoding (and its LLL-aided approximation) still achieve the maximum receive diversity, their gap with the optimal ML decoding grows unboundedly with SNR.

\section{System Model}

We consider a multiple-antenna system with \textit{M} transmit
antennas and \textit{N} receive antennas. In a multiple-access
system, we consider different transmit antennas as different
users. If we consider $ \mathbf{y}=[y_{1},...,y_{N}]^{T} $, $
\mathbf{x}=[x_{1},...,x_{M}]^{T} $, $
\mathbf{w}=[w_{1},...,w_{N}]^{T} $ and the $ N\times M $ matrix
$\mathbf{H}$, as the received signal, the transmitted signal, the
noise vector and the channel matrix, respectively, we have the
following matrix equation:
\begin{equation}
\mathbf{y}=\mathbf{H}\mathbf{x}+\mathbf{w} .
\end{equation}
The channel is assumed to be Raleigh,
i.e. the elements of $\mathbf{H}$ are i.i.d with the zero-mean
unit-variance complex Gaussian distribution, and the noise is Gaussian. Also, we have the
power constraint on the transmitted signal, $ \textmd{E} \Vert
\mathbf{x} \Vert^{2} =P$. The power of the additive noise is $
\sigma^{2} $ per antenna, i.e. $ \textmd{E} \Vert \mathbf{w}
\Vert^{2} =N\sigma^{2} $. The signal to noise ratio
(SNR) is defined as $ \rho=\frac{MP}{\sigma^{2}} $.

We send space-time codewords $\mathbf{X}=\left[ \mathbf{x}_{1},...,\mathbf{x}_{T}\right] $ with complex entries
($ \mathbf{x}_{i} \in \mathbb{C}^{M} $) and at the receiver, we find $
\tilde{\mathbf{x}}_{i} $ as $\mathbf{H}^{-1}\tilde{\mathbf{y}}_{i} $ where
$\left[ \tilde{\mathbf{y}}_{1},...,\tilde{\mathbf{y}}_{T}\right]$ is the closest $ MT$-dimensional lattice point to $\left[ \mathbf{y}_{1},...,\mathbf{y}_{T}\right]
$.

\section{Rate-diversity trade-off for the naive lattice decoding}

To drive the upper bound on the rate-diversity trade-off of NLD, we first present a lower bound on the probability that the received lattice (the lattice code after passing through the fading channel) has a short vector.

\begin{lemma}
Assume that the entries of the $ N\times M $ matrix $\mathbf{H}$
has independent complex Gaussian distributions with zero mean and
unit variance and consider $ d\left( \mathbf{H}_{T}\mathbf{L}\right)  $ as the minimum distance of
the lattice generated by $
\mathbf{H}_{T}\mathbf{L}$, where $ \mathbf{L}$ is the full-rank $
MT\times MT $ generator of a given complex lattice with unit volume\footnote{Volume of a lattice generated by matrix $ \mathbf{L}$ is defined as $\det \Lambda \triangleq \det (\mathbf{L}^{\ast}\mathbf{L})^{\frac{1}{2}}$, and is equal to the volume of the fundamental region of the lattice.} and $\mathbf{H}_{T}$ is the $NT \times MT $ block diagonal matrix constructed by repeating $ \mathbf{H}$ along the main diagonal. We have, 
\begin{equation} \lim_{\varepsilon \rightarrow 0} \frac{\log \Pr\lbrace  d\left( \mathbf{H}_{T}\mathbf{L}\right) \leq \varepsilon \rbrace}{\log \varepsilon}\leq 2M(N-M+1)
\end{equation}
 \end{lemma}

\begin{Proof} 
Consider $\sigma_{1}\leq \sigma_{2} \leq ...\leq \sigma_{M}$ the nonzero singular values of $\mathbf{H}$. 
Considering the pdf of the singular values of a Gaussian matrix \cite{Edelman}, it can be shown that \cite{ZT2003}

\begin{equation} \lim_{\varepsilon \rightarrow 0} \frac{\log \Pr\left\lbrace \sigma_{1}\leq \varepsilon^{b_{1}},..., \sigma_{M}\leq \varepsilon^{b_{M}}  \right\rbrace }{\log \varepsilon}= \sum_{i=1}^{M} 2(N-M+2i-1)b_{i}
\end{equation}

Thus

$$ \lim_{\varepsilon \rightarrow 0} \frac{\log \Pr\left\lbrace \sigma_{1}\leq \frac{1}{4\sqrt{M}}\varepsilon^{M},\sigma_{i}\leq \frac{1}{4\sqrt{M}}\;for \; i>1  \right\rbrace }{\log \varepsilon}= $$  $$ 2(N-M+1)\cdot \left( M+\lim_{\varepsilon \rightarrow 0} \frac{\log\frac{1}{4\sqrt{M}}}{\log \varepsilon}\right) +\sum_{i=2}^{M}2(N-M+2i-1)\cdot \lim_{\varepsilon \rightarrow 0} \frac{\log\frac{1}{4\sqrt{M}}}{\log \varepsilon}$$

\begin{equation} =  2M(N-M+1).
\end{equation}

Consider $\mathbf{v}_{min}$ as the singular vector of $\mathbf{H}$, corresponding to $\sigma_{1} $. For each $MT$-dimensional complex vector $\mathbf{v}=[a_{1}\mathbf{v}_{min}^{\mathsf{T}} \;\;a_{2} \mathbf{v}_{min}^{\mathsf{T}} ... \; a_{T}\mathbf{v}_{min}^{\mathsf{T}}]^{\mathsf{T}}$, 

\begin{equation}\Vert \mathbf{H}_{T}\mathbf{v}\Vert^{2} =\sum_{i=1}^{T} a_{i}^{2}\Vert \mathbf{H}\mathbf{v}_{min}\Vert^{2} =\sum_{i=1}^{T} \sigma_{1}^{2} \Vert a_{i}\mathbf{v}_{min}\Vert^{2} = \sigma_{1}^{2}\Vert \mathbf{v} \Vert^{2}.   
\end{equation}
Thus, assuming $ \sigma_{1} \leq \frac{1}{4\sqrt{M}}\varepsilon^{M}$,
\begin{equation} \Vert \mathbf{H}_{T}\mathbf{v}\Vert \leq \frac{1}{4\sqrt{M}}\varepsilon^{M}\Vert \mathbf{v} \Vert.
\end{equation}

Consider $ \mathcal{A}$ as a $2MT$-dimensional hypercube with edges of length $ \frac{1}{ \varepsilon^{M} } $ whose $2T$ edges are parallel to the subspace spanned by the vectors $\mathbf{v}=[a_{1}\mathbf{v}_{min}^{\mathsf{T}} \;\;a_{2} \mathbf{v}_{min}^{\mathsf{T}} ... \; a_{T}\mathbf{v}_{min}^{\mathsf{T}}]^{\mathsf{T}}$ and the other $2T(M-1)$ edges are orthogonal to that subspace. The volume of this cube is $\varepsilon^{-2M^{2}T}$. Because the volume of the lattice is 1, for $K$, the number of lattice points inside this cube, we have\footnote{When a region is large, the number of lattice points inside the region can be approximated by the ratio between the volume of the region and the volume of the lattice.} $\lim_{\varepsilon \rightarrow 0} \frac{K}{\varepsilon^{-2M^{2}T}}=1 $.

Now, assuming $\sigma_{1}\leq \frac{1}{4\sqrt{M}}\varepsilon^{M}$ and $\sigma_{M}\leq \frac{1}{4\sqrt{M}}$, the region $ \mathbf{H}_{T}\mathcal{A}$ is inside a $2MT$-dimensional orthotope (in the subspace spanned by $ \mathbf{H}_{T} $) whose $2T $ edges (which correspond to the smallest singular value $ \sigma_{1}$) have length $\frac{1}{4\sqrt{M}}$ and the length of the other $2T(M-1) $ is at most $  \frac{1}{4\sqrt{M}\varepsilon^{M}}$ (because of the bound on the largest singular value $ \sigma_{M}$). The $ 2T$ smaller edges can be covered by at most $\lceil 4^{-1}\varepsilon^{-1}\rceil \leq 2^{-1}\varepsilon^{-1} $ segments of length $ \frac{\varepsilon}{\sqrt{M}}$ and the others can be covered by at most $ \lceil 4^{-1}\varepsilon^{-(M+1)}\rceil \leq 2^{-1}\varepsilon^{-(M+1)} $ segments of length $ \frac{\varepsilon}{\sqrt{M}}$. Thus, this orthotope can be covered by at most $\left( 2^{-1}\varepsilon^{-1}\right)^{2T}\left( 2^{-1}\varepsilon^{-(M+1)}\right)^{2T(M-1)} = 2^{-2MT}\varepsilon^{-2M^{2}T} $ hypercubes of edge length $\frac{\varepsilon}{\sqrt{M}}$. Because $\lim_{\varepsilon \rightarrow 0} \frac{K}{\varepsilon^{-2M^{2}T}}=1 $, when $\varepsilon \rightarrow 0$, the number of these small hypercubes is smaller than the number of lattice points inside them. Thus, based on Dirichlet's box principle, in one of these hypercubes there are at least 2 points of the new lattice, hence $ d\left( \mathbf{H}_{T}\mathbf{L}\right)$ is smaller than the diameter of the small hyper cubes:
\begin{equation}d_{\mathbf{H}}\leq \sqrt{M}\cdot \frac{\varepsilon}{\sqrt{M}}. 
\end{equation}

Therefore, 

\begin{equation}\lim_{\varepsilon \rightarrow 0} \frac{\log \Pr\lbrace  d\left( \mathbf{H}_{T}\mathbf{L}\right) \leq \varepsilon \rbrace}{\log \varepsilon}\leq \lim_{\varepsilon \rightarrow 0} \frac{\log \Pr\left\lbrace \sigma_{1}\leq \varepsilon^{M},\sigma_{M}\leq \frac{1}{2M}  \right\rbrace }{\log \varepsilon}=2M(N-M+1). 
\end{equation}

\end{Proof}

\begin{thm}
Consider a MIMO fading channel with $ M $ transmit and $ N $
receive antennas ($M\leq N$) with codebooks from an $MT$-dimensional lattice $ \mathbf{L}$, which are sent over $T$ channel uses. For the naive lattice decoding, the rate-diversity trade-off of the system is 

$$ d_{NLD}(r)\leq M(N-M+1)-r \left(N-M+1\right) ,$$
\begin{equation}\;\;\;\;\;\;\;\;\;\;\;\;\;\;\;\;\;\; \;\;\;\;\;\;\;\;\; \;\;\;\;\;\;\;\;\; {\rm for} \; 0 \leq r\leq M.  
\end{equation}
 \end{thm}

\begin{Proof}
Consider the code of rate $R$ constructed from the lattice. The number of codewords is equal to $2^{R} $. Without any loss of generality, we can assume that the volume of the lattice is fixed and is equal to 1, and the power constraint $ P$ is dependent on the rate. To satisfy the power constraint, at least half of the codewords should have power less than $ 2P$. The number of codewords with power less than $2P$ is equal to the number of lattice points inside a $2M$-dimensional sphere whose volume is proportional to $ P^{M}$. Thus, by approximating the number of lattice points with the ratio of the volume of the region and the volume of the lattice:

\begin{equation} \label{eq:6}
2^{R} \leq c_{1}P^{M}.
\end{equation}
where $ c_{1}$ is a constant, independent of SNR\footnote{Throughout this paper $c_{1},c_{2},...$ are only dependent on size of dimensions.}. 
According to the definition of the multiplexing gain, $ r=\lim_{SNR\rightarrow \infty} \frac{\log R}{\log SNR}$. Using (\ref{eq:6}), 

\begin{equation} \label{eq:7}
\lim_{SNR\rightarrow \infty} \frac{\log P}{\log SNR} \geq \frac{\log \frac{1}{M} \log R }{\log SNR} =\frac{r}{M} .
\end{equation}

For the symbol error probability $P_{e}$, considering $ SNR = \frac{MP}{\sigma^{2}} $,

\begin{equation}P_{e} \geq \Pr \left\lbrace d\left( \mathbf{H}_{T}\mathbf{L}\right)\leq  \frac{\sigma}{\sqrt{M}} \right\rbrace . Q\left (\frac{1}{2\sqrt{M}} \right) = \Pr \left\lbrace d\left( \mathbf{H}_{T}\mathbf{L}\right)\leq  \frac{\sqrt{P}}{\sqrt{SNR}}\right\rbrace . Q\left (\frac{1}{2\sqrt{M}} \right).
\end{equation}

Therefore, using lemma 1 (with $\varepsilon= \frac{\sqrt{P}}{\sqrt{SNR}}$) and (\ref{eq:7}),
$$ d_{NLD}(r) = \lim_{SNR\rightarrow \infty} \frac{-\log P_{e}}{\log SNR} \leq \lim_{SNR\rightarrow \infty} \frac{-\log  \Pr \left\lbrace d_{\mathbf{H}}\leq  \frac{\sqrt{P}}{\sqrt{SNR}}\right\rbrace}{\log SNR} $$

$$\leq \lim_{SNR\rightarrow \infty} \frac{-2M(N-M+1) \left( \log  \frac{\sqrt{P}}{\sqrt{SNR}} \right) }{\log SNR}  $$

$$= \lim_{SNR\rightarrow \infty} \frac{-2M(N-M+1) \left( \frac{1}{2}\log  P - \frac{1}{2} \log SNR \right) }{\log SNR}  $$

$$ \leq -\left( \frac{r}{2M}-\frac{1}{2}\right)\cdot 2M(N-M+1) $$

\begin{equation} =M(N-M+1)-r \left(N-M+1\right).
\end{equation}
\end{Proof}

\begin{cor}
In a MIMO fading channel with $ M = N $ transmit and receive antennas, if we use the naive lattice decoding, the rate-diversity trade-off for full-rate lattice code can not be better than that of V-BLAST.
\end{cor}

\begin{Proof}
When $M=N$, according to Theorem 1,

\begin{equation} d_{NLD}(r) \leq M-r
\end{equation}
On the other hand, for the V-BLAST system with lattice decoding \cite{Tse-book},
\begin{equation} d_{V-BLAST}(r) = M-r
\end{equation}
\end{Proof}

It is interesting to compare this result with the results on lattice space-time codes which have non-vanishing determinants. Although by ML decoding, these codes (such as the $2\times 2$ Golden code) achieve the optimal rate-diversity trade-off, when we replace ML decoding with the naive lattice decoding (and its approximations), their performance is not much better than the simple V-BLAST scheme (specially when the number of transmit and receive antennas are the same) 

\begin{figure}
  \centering
  \includegraphics[scale=1,clip]{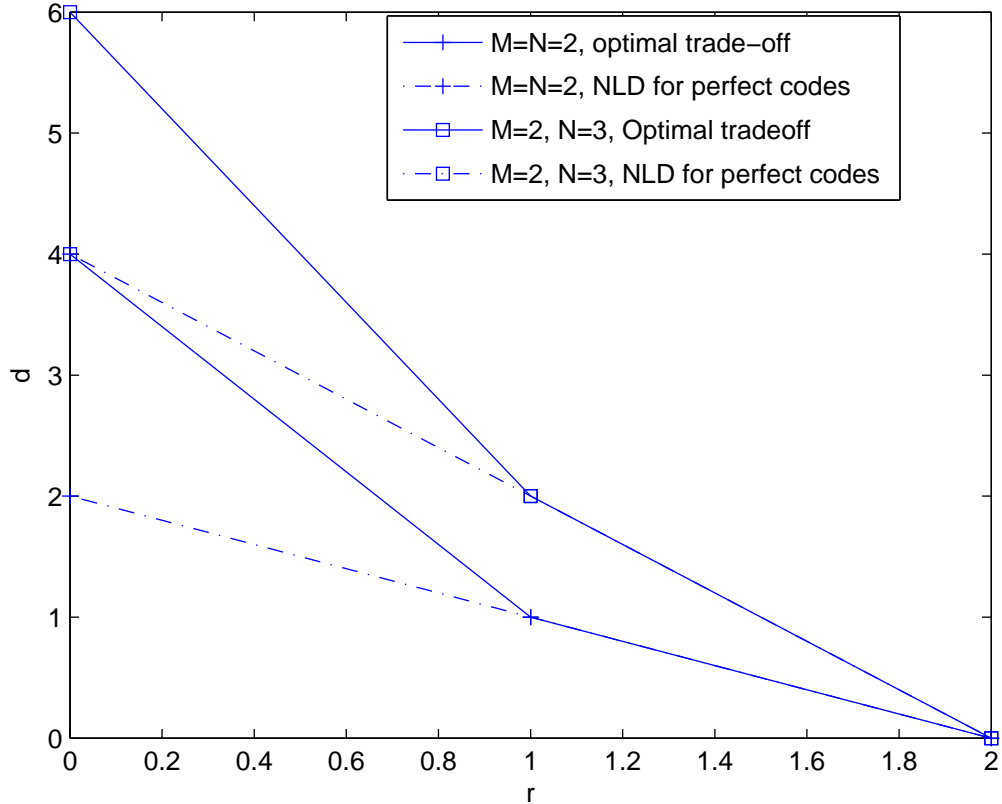}

  \caption{Comparison between the optimal rate-diversity tradeoff and the upper bound on the rate-diversity trade-off of full-rate lattice codes (including perfect space-time codes such as the Golden code)}
  \label{}
\end{figure}

To better understand the difference between the naive lattice decoding and the ML decoding, we note that for small constellations, when the generator of the received lattice has a small singular value, the minimum distance of the lattice can be much smaller than the minimum distance of the constellation. Figure 2 shows this situation for a small 4-point constellation from a 2-dimensional lattice.

We should note that this upper bound is for full-rate lattices. Lattices with lower rate, can provide higher diversity, but their rate is limited by the dimension of the lattice. For example, The Alamouti code, based on QAM constellations, can achieve the full diversity for fixed rates ($r=0$), but its rate is limited by one.

\begin{figure}
  \centering
  \includegraphics[scale=.8,clip]{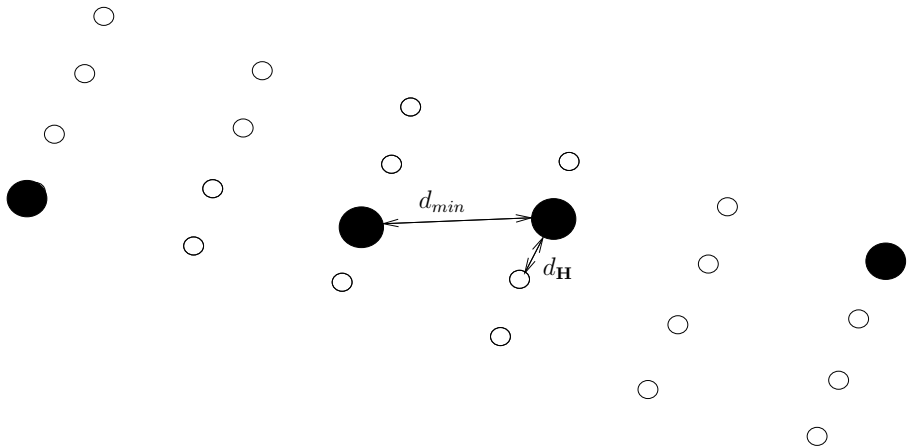}

  \caption{Minimum distance of a lattice ($d_{\mathbf{H}}=d\left(\mathbf{H}_{T} \mathbf{L} \right) $), compared to the minimum distance of a lattice code ($d_{min}$)}
  \label{}
\end{figure}

\section{Asymptotic performance of the naive lattice decoding for $M=N$}

In \cite{taherzad-IT2}, it is shown that for $N\geq M$, the naive lattice decoding achieves the receive diversity in V-BLAST systems (indeed, even its simple latice-reduction-aided approximation still achieves the optimum receive diversity of order $N$). However, there is a difference between two cases of $M<N$ and $M=N $. While for $M<N$, compared to ML decoding, the performance loss of the naive lattice decoding is bounded in terms of SNR  \cite{taherzad-IT2}, here we show this is not valid for the case of $M=N$. This dichotomy is related to the bounds on the probability of having a short lattice vector in a lattice generated by a random Gaussian matrix. 

In \cite{taherzad-IT1}, an upper bound on the probability of having a short lattice vector is given:

\begin{lemma}
Assume that the entries of the $ M\times M $ matrix $\mathbf{H}$
has independent complex Gaussian distributions with zero mean and
unit variance and consider $ d(\mathbf{H}) $ as the minimum distance of
the lattice generated by $ \mathbf{H} $.
 Then, there is a constant $ C $ such that \cite{taherzad-IT1},
$$ Prob\lbrace d(\mathbf{H})\leq \varepsilon \rbrace \leq C \varepsilon^{2M}\ln \left( \frac{1}{\varepsilon}\right)^{M-1} .  $$
 \end{lemma}

The term $\ln \left( \frac{1}{\varepsilon}\right)$ suggests an unboundedly increasing gap between the performance of ML decoding and the naive lattice decoding (though both of them have the same slope $M$). 

In this section, we present a lower bound on the error probability of the naive lattice decoding and show that this unboundedly increasing gap does exist. 
\begin{lemma}
For $M \geq 2 $ and $\varepsilon <1 $, for the lattice generated by an $ M\times M $ random complex Gaussian matrix $\mathbf{H}$ with zero mean and
unit variance, there is a constant $ C^{\prime} $ such that,
\begin{equation} Prob\lbrace d(\mathbf{H})\leq \varepsilon \rbrace \geq C^{\prime} \varepsilon^{2M}\ln \left( \frac{1}{\varepsilon}\right) .  
\end{equation}
 \end{lemma}

\begin{Proof}
Consider $ L_{(\mathbf{v}_{1},...,\mathbf{v}_{M})} $ as the lattice generated by $\mathbf{v}_{1} $,$\mathbf{v}_{2} $,...,$\mathbf{v}_{M}$. Each point of $ L_{(\mathbf{v}_{1},...,\mathbf{v}_{M})} $ can
be represented by $ \mathbf{v}_{(z_{1},...,z_{M})}=z_{1}\mathbf{v}_{1} + z_{2}\mathbf{v}_{2}+ ...+z_{M}\mathbf{v}_{M}$, where $ z_{1},...,z_{M}$ are complex integer numbers. 

The vectors $\mathbf{v}_{1} $,$\mathbf{v}_{2} $,...,$\mathbf{v}_{M}$ are independent and jointly Gaussian. Therefore, for every complex vector $ \mathbf{b}=(b_{1},...,b_{M})$, the vector $\mathbf{v}_{\mathbf{b}}=b_{1}\mathbf{v}_{1} + b_{2}\mathbf{v}_{2}+ ...+b_{M}\mathbf{v}_{M}$ has complex circular Gaussian distribution with the variance 

\begin{equation}
\varrho_{\mathbf{b}}^{2}=\Vert \mathbf{b} \Vert^{2} = \vert b_{1}\vert^{2}+...+\vert b_{M}\vert^{2}.
\end{equation}

Now, considering the pdf of $\mathbf{v}_{\mathbf{b}}$, we can bound $\Pr\left\lbrace \Vert \mathbf{v}_{\mathbf{b}} \Vert \leq \varepsilon \right\rbrace = \int_{\Vert \mathbf{v} \Vert \leq \varepsilon} f_{\mathbf{v}_{\mathbf{b}}}(\mathbf{v}) \; d\mathbf{v}$ by using the fact that $e^{-\frac{\varepsilon^{2}}{\varrho_{\mathbf{b}}^{2}}} \leq   e^{-\frac{\Vert \mathbf{v} \Vert^{2}}{\varrho_{\mathbf{b}}^{2}}} \leq 1$ for $ \Vert \mathbf{v} \Vert \leq \varepsilon$:

\begin{equation} \int_{\Vert \mathbf{v} \Vert \leq \varepsilon} \frac{1}{\pi^{M} \varrho_{\mathbf{b}}^{2M}} e^{-\frac{\varepsilon^{2}}{\varrho_{\mathbf{b}}^{2}}}\; d\mathbf{v} \leq  \int_{\Vert \mathbf{v} \Vert \leq \varepsilon} f_{\mathbf{v}}(\mathbf{v}) \; d\mathbf{v} \leq \int_{\Vert \mathbf{v} \Vert \leq \varepsilon} \frac{1}{\pi^{M} \varrho_{\mathbf{b}}^{2M}} \; d\mathbf{v}. \end{equation}

Thus, because the volume of region of the integral (which is a $2M$-dimensional sphere with radius $\varepsilon$) is proportional to $\varepsilon^{2M} $,

\begin{equation} \label{eq:18}
c_{6}\frac{\varepsilon^{2M}}{\Vert \mathbf{b} \Vert^{2M} }e^{-\frac{\varepsilon^{2}}{\Vert \mathbf{b} \Vert^{2}}}\leq \Pr\left\lbrace \Vert \mathbf{v}_{\mathbf{b}} \Vert \leq \varepsilon\right\rbrace  \leq c_{7}\frac{\varepsilon^{2M}}{\Vert \mathbf{b} \Vert^{2M} }. 
\end{equation}

We can represent any $ M$-dimensional complex integer vector as a $ 2M$-dimensional real integer vector. In our proof, we consider only integer vectors in the set $\mathcal{B} $ which consists of integer vectors $\mathbf{z} $ such that their real entries do not have a nontrivial common divisor and $\Vert \mathbf{z}\Vert_{\infty} \leq \varepsilon^{-\frac{1}{2M}}$ where $\Vert \cdot \Vert_{\infty}$ represents the norm of the largest real entry. First, we show that the number of such integer vectors $\mathbf{z}$ in the region $2^{(k-1)} <\Vert\mathbf{z}\Vert_{\infty} \leq 2^k$ is at least $2^{2Mk}$.
The total number of integer points in the region $2^{(k-1)} <\Vert\mathbf{z}\Vert_{\infty} \leq 2^k$ is\footnote{The number of points in the cube $\Vert\mathbf{z}\Vert_{\infty} \leq 2^k$ is $\left( 2^{k+1}+1\right)^{2M}$ and the number of points in the cube $\Vert\mathbf{z}\Vert_{\infty}\leq 2^{(k-1)} $ is $\left( 2^{k}+1\right)^{2M}$.} $\left( 2^{k+1}+1\right)^{2M} - \left( 2^{k}+1\right)^{2M}$. The number of those points whose entries have a common divisor $i $ is at most equal to the number of integer points in the region $ \Vert\mathbf{z}\Vert_{\infty} \leq \frac{2^k}{i}$. Therefore, $n_{k} $, the number of integer vectors $\mathbf{z}$ whose entries does not have nontrivial common divisors, can be lower bounded by

$$n_{k} \geq  \left( \left( 2^{k+1}+1\right)^{2M} - \left( 2^{k}+1\right)^{2M}\right) - \sum_{i=2}^{2^{k}}\left( 2\frac{2^{k}}{i}+1\right)^{2M}$$
$$ >\left( \left( 2^{k+1}+1\right)^{2M} - \left( 3\cdot 2^{k-1}\right)^{2M}\right) - \sum_{i=2}^{2^{k}}\left( 3\frac{2^{k}}{i}\right)^{2M}$$
$$ > 2^{2kM+2M} \left( 1-\left( \frac{3}{4}\right)^{2M}  - \left( \frac{3}{2}\right)^{2M}\sum_{i=2}^{\infty} \frac{1}{i^{2M}}\right) $$
$$ > 2^{2kM+2M} \left( 1-\left( \frac{3}{4}\right)^{2M}  - \left( \frac{3}{2}\right)^{2M}\cdot \left( \frac{1}{2^{2M}}+ \frac{1}{3^{2M}}+ \int_{3}^{\infty} \frac{1}{x^{2M}}\; dx\right)\right)  $$
$$=  2^{2kM+2M} \left( 1-\left( \frac{3}{4}\right)^{2M}  -  \left( \frac{3}{4}\right)^{2M} - \left( \frac{1}{2}\right)^{2M}- \left( \frac{3}{2}\right)^{2M}\cdot \frac{1}{3^{2M-1}(2M-1)}\right) $$
$$ > 2^{2kM+2M} \left( 1-2\left( \frac{3}{4}\right)^{2M} -\frac{1}{2^{2M}}.\left( 1+ \frac{2}{2M-1}\right) \right)  $$
\begin{equation} \label{eq:19}
\geq  2^{2kM+2M} \left( 1-2\left( \frac{3}{4}\right)^{4} -\frac{1}{2^{4}}.\left( 1+ \frac{2}{3}\right) \right)  >2^{2kM+2M}\cdot 2^{-4}  \geq 2^{2kM} \;\;\;\; {\rm for} \; M\geq 2.  
\end{equation}

\begin{figure}
  \centering
  \includegraphics[scale=.6,clip]{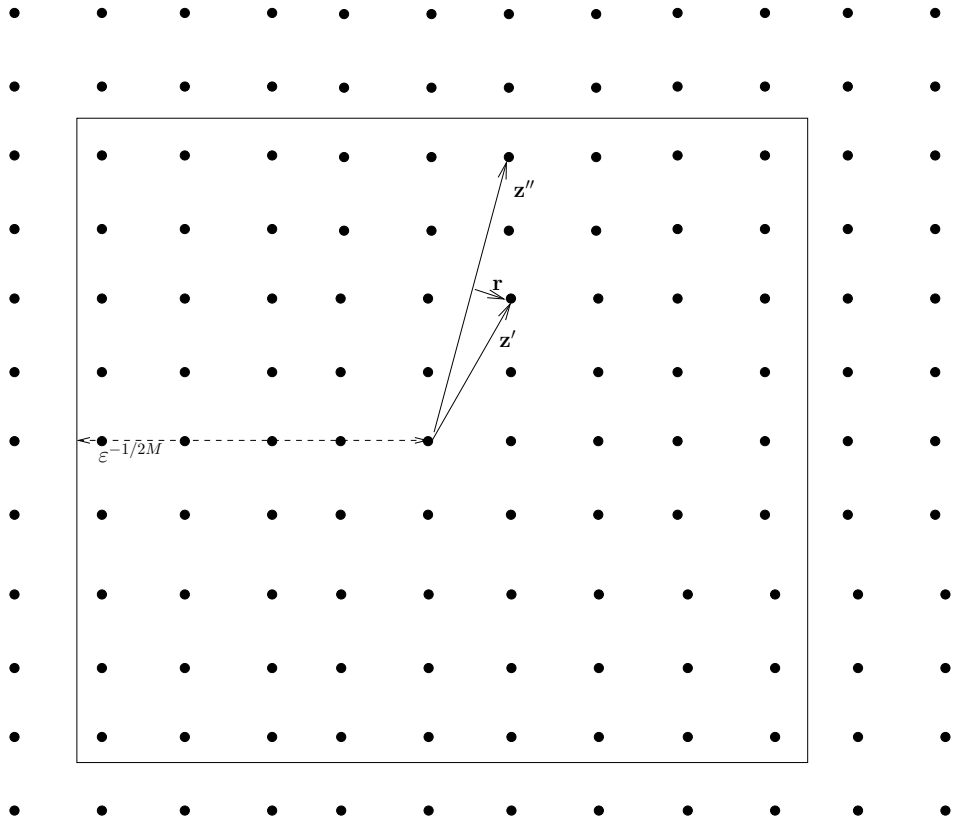}

  \caption{integer points in the region $ \Vert\mathbf{z}\Vert_{\infty} \leq \varepsilon^{-\frac{1}{2M}}$.}
  \label{}
\end{figure}

Now, we find an upper bound on $\Pr\left\lbrace \Vert \mathbf{v}_{\mathbf{z}^{\prime}} \Vert \leq \varepsilon, \Vert \mathbf{v}_{\mathbf{z}^{\prime \prime}} \Vert \leq \varepsilon \right\rbrace$ for two different complex integer vectors $\mathbf{z}^{\prime}$ and $\mathbf{z}^{\prime \prime} $ which belong to  $ \mathcal{B}$. We can write $\mathbf{z}^{\prime}$ as $a \mathbf{z}^{\prime \prime}+  \mathbf{r}$ where $a$ is a complex number and $ \mathbf{r}$ is a complex vector, orthogonal to $\mathbf{z}^{\prime \prime} $. We show that $\Vert \mathbf{r} \Vert \geq \frac{1}{\sqrt{2M}} \varepsilon^{\frac{1}{2M}}$.
The area of the triangle which has vertexes $\mathbf{0}$, $\mathbf{z}^{\prime}$, and $\mathbf{z}^{\prime \prime}$, is equal to $S=\frac{1}{2}\Vert \mathbf{r} \Vert \cdot \Vert \mathbf{z}^{\prime \prime} \Vert $. On the other hand, because $\mathbf{0}$, $\mathbf{z}^{\prime}$, and $\mathbf{z}^{\prime \prime}$ are integer points, $ 2S$ should be integer. Also, because the entries of $\mathbf{z}^{\prime}$ do not have any nontrivial common divisor, $\mathbf{z}^{\prime}$ can not be a multiplier of $\mathbf{z}^{\prime \prime}$ (and vice versa). Because $\mathbf{z}^{\prime}$ and $\mathbf{z}^{\prime \prime}$ are not multipliers of each other, $ S$ is nonzero. Thus, $ S\geq \frac{1}{2}$, hence,

\begin{equation}
\frac{1}{2}\Vert \mathbf{r} \Vert \cdot \Vert \mathbf{z}^{\prime \prime} \Vert \geq \frac{1}{2}
\end{equation}

\begin{equation} \label{eq:21}
\Longrightarrow \Vert \mathbf{r} \Vert \geq\frac{1}{\Vert \mathbf{z}^{\prime \prime} \Vert} \geq \frac{1}{\sqrt{2M}\Vert \mathbf{z}^{\prime \prime} \Vert_{\infty}} \geq \frac{1}{\sqrt{2M} \varepsilon^{-\frac{1}{2M}}} =\frac{1}{\sqrt{2M}}\varepsilon^{\frac{1}{2M}}.
\end{equation}

Now we bound $ \Pr\left\lbrace \Vert \mathbf{v}_{\mathbf{z}^{\prime}} \Vert \leq \varepsilon, \Vert \mathbf{v}_{\mathbf{z}^{\prime \prime}} \Vert \leq \varepsilon \right\rbrace$. Because $ \mathbf{r} \perp \mathbf{z}^{\prime \prime}  $, we can see that $\mathbf{v}_{\mathbf{r}} \perp \mathbf{v}_{ \mathbf{z}^{\prime \prime}} $. Thus, when $\Vert \mathbf{v}_{\mathbf{r}} \Vert > \varepsilon$, using the fact that $ \mathbf{v}_{\mathbf{a}+\mathbf{b}}=\mathbf{v}_{\mathbf{a}}+\mathbf{v}_{\mathbf{b}}$, 

$$ \Vert \mathbf{v}_{\mathbf{z}^{\prime}} \Vert = \Vert \mathbf{v}_{a \mathbf{z}^{\prime \prime} + \mathbf{r}} \Vert = \Vert \mathbf{v}_{a \mathbf{z}^{\prime \prime}} + \mathbf{v}_{\mathbf{r}} \Vert \geq \Vert \mathbf{v}_{\mathbf{r}} \Vert > \varepsilon$$
Therefore, 
$$  \Pr\left\lbrace \Vert \mathbf{v}_{\mathbf{z}^{\prime}} \Vert \leq \varepsilon, \Vert \mathbf{v}_{\mathbf{z}^{\prime \prime}} \Vert \leq \varepsilon \right\rbrace
$$
\begin{equation} \label{eq:r-z} \leq \Pr\left\lbrace \Vert \mathbf{v}_{\mathbf{r}} \Vert \leq \varepsilon, \Vert \mathbf{v}_{\mathbf{z}^{\prime \prime}} \Vert \leq \varepsilon \right\rbrace
\end{equation}
Based on the orthogonality of $ \mathbf{r}$ and $\mathbf{z}^{\prime \prime} $, $\mathbf{v}_{\mathbf{r}} $ and $\mathbf{v}_{\mathbf{\mathbf{z}^{\prime \prime}}}$ are independent. Thus, using (\ref{eq:18}), (\ref{eq:21}), and noting that $\Vert \mathbf{z}^{\prime \prime} \Vert \geq 1$ (because $ \mathbf{z}^{\prime \prime}$ is a nonzero integer vector):
$$\Pr\left\lbrace \Vert \mathbf{v}_{\mathbf{z}^{\prime}} \Vert \leq \varepsilon, \Vert \mathbf{v}_{\mathbf{z}^{\prime \prime}} \Vert \leq \varepsilon \right\rbrace \leq \Pr\left\lbrace \Vert \mathbf{v}_{\mathbf{z}^{\prime \prime}} \Vert \leq \varepsilon \right\rbrace \cdot \Pr\left\lbrace \Vert \mathbf{v}_{\mathbf{r}} \Vert \leq \varepsilon \right\rbrace$$

$$ \leq \left(  c_{7} \frac{\varepsilon^{2M}}{\Vert \mathbf{z}^{\prime \prime} \Vert^{2M}}\right) \cdot \left(  c_{7} \frac{\varepsilon^{2M} }{\Vert \mathbf{r} \Vert^{2M}}\right) $$

$$ \leq c_{7}^{2} \varepsilon^{2M} \cdot \frac{\varepsilon^{2M}\left(2M \right)^{M}}{\left( \varepsilon^{\frac{1}{2M}}\right) ^{2M}}$$

\begin{equation} \label{eq:double} = c_{8} \varepsilon^{4M-1} 
\end{equation}

Now, we use the Bonferroni inequality \cite{Bonferroni},

$$\Pr\left\lbrace d(\mathbf{H}) \leq \varepsilon\right\rbrace =  \Pr\left\lbrace \exists \; \mathbf{z} \neq 0:\; \Vert \mathbf{v}_{\mathbf{z}} \Vert \leq \varepsilon \right\rbrace  \geq \Pr\left\lbrace \exists \;\mathbf{z}: \;  \mathbf{z} \in \mathcal{B}, \Vert \mathbf{v}_{\mathbf{z}} \Vert \leq \varepsilon \right\rbrace$$

$$ \geq \sum_{\mathbf{z} \in \mathcal{B}}\Pr\left\lbrace \Vert \mathbf{v}_{\mathbf{z}} \Vert \leq \varepsilon\right\rbrace  
$$

\begin{equation} \label{eq:bonf} -\sum_{ \mathbf{z}^{\prime}, \mathbf{z}^{\prime\prime} \in \mathcal{B}}   \Pr\left\lbrace \Vert \mathbf{v}_{\mathbf{z}^{\prime}} \Vert \leq \varepsilon, \Vert \mathbf{v}_{\mathbf{z}^{\prime \prime}} \Vert \leq \varepsilon \right\rbrace
\end{equation}

For the first term of (\ref{eq:bonf}), 

\begin{equation}  \sum_{\mathbf{z} \in \mathcal{B}}\Pr\left\lbrace \Vert \mathbf{v}_{\mathbf{z}} \Vert \leq \varepsilon\right\rbrace  
\end{equation}

\begin{equation} \label{eq:26}
  \geq \sum_{k=0}^{\left \lfloor \log\left( \varepsilon^{-\frac{1}{2M}}\right) \right \rfloor }   \sum_{\mathbf{z} \in \mathcal{B},2^{k-1}<\Vert \mathbf{z} \Vert_{\infty} \leq 2^{k}}\Pr\left\lbrace \Vert \mathbf{v}_{\mathbf{z}} \Vert \leq \varepsilon\right\rbrace  
\end{equation}
By using (\ref{eq:19}), (\ref{eq:18}), and noting that $\Vert \mathbf{z} \Vert \leq  \sqrt{2M}\Vert \mathbf{z} \Vert_{\infty}$ and $e^{-\frac{\varepsilon^{2}}{\Vert \mathbf{z} \Vert^{2}}} \geq e^{-1} $ (because $\varepsilon <1$ and $\Vert \mathbf{z} \Vert \geq 1 $),

\begin{equation}
(\ref{eq:26}) \geq \sum_{k=0}^{\left \lfloor \log\left( \varepsilon^{-\frac{1}{2M}}\right) \right \rfloor  }2^{2kM}\cdot \frac{c_{6}\varepsilon^{2M}}{\left( 2^k\right) ^{2M}\cdot (2M)^{M}} \cdot e^{-1}
\end{equation}

\begin{equation}
\geq \sum_{k=0}^{\left \lfloor \log\left( \varepsilon^{-\frac{1}{2M}}\right) \right \rfloor } c_{9}\varepsilon^{2M}
\end{equation}

\begin{equation} \label{eq:term1}
=\left( \left \lfloor \log\left( \varepsilon^{-\frac{1}{2M}}\right) \right \rfloor+1\right)\cdot c_{9}\varepsilon^{2M} 
\geq c_{10}\varepsilon^{2M}\cdot \ln \left(\frac{1}{\varepsilon} \right). 
\end{equation}

For the second term of of (\ref{eq:bonf}), because the number of complex integers in $\mathcal{B} $ (which is at most the number of integer points in the cube $ \Vert \mathbf{z} \Vert_{\infty} \leq \varepsilon^{-\frac{1}{2M}}$) is bounded by $c_{11} \left(\varepsilon^{-\frac{1}{2M}} \right) ^{2M}=c_{11} \varepsilon^{-1} $, the number of pairs $(\mathbf{z}^{\prime},\mathbf{z}^{\prime\prime})$ is at most $\left( c_{11} \varepsilon^{-1}\right)^{2}$. Thus, using (\ref{eq:double}):

\begin{equation}  \sum_{\mathbf{z}^{\prime},\mathbf{z}^{\prime\prime}\in \mathcal{B}}\Pr\left\lbrace \Vert \mathbf{v}_{\mathbf{z}^{\prime}} \Vert \leq \varepsilon, \Vert \mathbf{v}_{\mathbf{z}^{\prime \prime}} \Vert \leq \varepsilon \right\rbrace
\end{equation}

\begin{equation}  \leq \left( c_{11} \varepsilon^{-1}\right)^{2} \cdot c_{8} \varepsilon^{4M-1}
\end{equation}

\begin{equation} \label{eq:term2} \leq  c_{12} \varepsilon^{4M-3}
\end{equation}

Now, by using (\ref{eq:term1}) and (\ref{eq:term2}),

\begin{equation} (\ref{eq:bonf}) \geq  c_{10}\varepsilon^{2M} \ln \left(\frac{1}{\varepsilon} \right) - c_{12} \varepsilon^{4M-3}
\end{equation}

\begin{equation} \geq C^\prime \varepsilon^{2M} \ln \left(\frac{1}{\varepsilon} \right), \;\;\;{\rm for}\; M\geq 2.
\end{equation}
\end{Proof}

\begin{thm}
Consider a MIMO fading channel with $ M $ transmit and $ M $
receive antennas and a V-BLAST transmission system. The naive
lattice-decoding has an asymptotically unbounded loss, campared to
the exact ML decoding.
 \end{thm}

\begin{Proof}
For ML decoding, by using the Chernoff bound for the pairwise error probability and then applying the union bound for the finite constellation, we have \cite{Tarokh98}
\begin{equation} P_{error-ML} \leq c_{13} (SNR)^{-M}
\end{equation}
where $ c_{13}$ depends on the size of constellation.

For naive lattice decoding,

$$ P_{error-NLD} \geq \Pr \left\lbrace d_{\mathbf{H}}\leq  \frac{1}{\sqrt{SNR}}\right\rbrace . Q\left (\frac{1}{\sqrt{M}} \right) $$

\begin{equation} \geq c_{14} (SNR)^{-M}\ln (SNR). 
\end{equation}
Therefore, although both of them asymptotically have the same slope and achieve the optimal receive diversity of order $M$, for large SNRs, the gap between their performances is unbounded (with a logarithmic growth, or in other words, $\log \log SNR$ in dB scale).
\end{Proof}

\section{conclusions}
In this paper, the inherent limitations of the performance of the naive lattice decoding is investigated. The naive lattice decoding and various implementions of it (such as the sphere decoding) and its simple approximated versions (such as the LLL-aided decoding) are very attractive for the practical MIMO systems. Nontheless, to achieve theoretical benchmarks (such as the rate-diversity trade-off), these techniques are not always sufficient. For the rate-diversity trade-off, although different elegant lattice codes  have been introduced which achieve the optimal trade-off \cite{KumarRDT1} \cite{BelfioreRDT} \cite{KumarRDT2}, they need ML decoding to achieve optimality. On the other hand, there can exist a family of lattice codes (based on different lattice structures for different rates and SNR values) which achives the optimum tradeoff using the naive lattice decoding \cite{ElGamal2007}. However, the existence proof in \cite{ElGamal2007} does not provide any constructive solution for the encoding of such codes. Therefore, the problem of achieving the optimum diversity-multiplexing tradeoff by a practical encoding and decoding scheme is still open.

\newcommand{\noopsort}[1]{} \newcommand{\printfirst}[2]{#1}
  \newcommand{\singleletter}[1]{#1} \newcommand{\switchargs}[2]{#2#1}

\end{document}